\documentclass[conference]{IEEEtran}

\usepackage{mathptmx}
\usepackage{cite}
\usepackage{amssymb,amsmath,amsfonts,amsthm}
\usepackage{algorithmic}
\usepackage{graphicx}
\usepackage{textcomp}
\usepackage{xcolor}
\usepackage{soul}
\usepackage{fancyhdr}
\usepackage{caption}
\usepackage{comment}
\usepackage{multirow}
\usepackage{enumitem}
\usepackage{todonotes}
\usepackage{subfig}
\usepackage[ruled,vlined,linesnumbered,noend]{algorithm2e}

\usepackage[hyphens]{url}
\usepackage[breaklinks]{hyperref}
\usepackage{breakurl}

\newcommand{\ignore}[1]{}
\newtheorem*{thm*}{Theorem}

\newcommand{\Z}{\mathbb{Z}}

\newcommand{\mat}[1]{\mathbf{#1}}%

\def\BibTeX{{\rm B\kern-.05em{\sc i\kern-.025em b}\kern-.08em
    T\kern-.1667em\lower.7ex\hbox{E}\kern-.125emX}}

\SetCommentSty{mycommfont}

\newcommand{\AccName}{XCRYPT}

\title{XCRYPT: Accelerating Lattice Based Cryptography \\ with Memristor Crossbar Arrays}

\author{\IEEEauthorblockN{Sarabjeet Singh}
\IEEEauthorblockA{University of Utah \\
sarab@cs.utah.edu}
\and
\IEEEauthorblockN{Xiong Fan}
\IEEEauthorblockA{Rutgers University \\
leofanxiong@gmail.com}
\and
\IEEEauthorblockN{Ananth Krishna Prasad}
\IEEEauthorblockA{University of Utah \\
ananth@cs.utah.edu}
\and
\IEEEauthorblockN{Lin Jia}
\IEEEauthorblockA{University of Utah \\
lin.jia@utah.edu}
\and
\IEEEauthorblockN{Anirban Nag}
\IEEEauthorblockA{Uppsala University \\
anirban.nag@it.uu.se}
\and
\IEEEauthorblockN{Rajeev Balasubramonian}
\IEEEauthorblockA{University of Utah \\
rajeev@cs.utah.edu}
\and
\IEEEauthorblockN{Mahdi Nazm Bojnordi}
\IEEEauthorblockA{University of Utah \\
mnbojnordi@gmail.com}
\and
\IEEEauthorblockN{Elaine Shi}
\IEEEauthorblockA{Carnegie Mellon University \\
runting@gmail.com}
}

\begin{document}
\maketitle
\thispagestyle{plain}
\pagestyle{plain}


\begin{abstract}

This paper makes a case for accelerating lattice-based post quantum cryptography (PQC) with memristor based crossbars, and shows that these inherently error-tolerant algorithms are a good fit for noisy analog MAC operations in crossbars. We compare different NIST round-3 lattice-based candidates for PQC, and identify that SABER is not only a front-runner when executing on traditional systems, but it is also amenable to acceleration with crossbars.  SABER is a module-LWR based approach, which performs modular polynomial multiplications with rounding. We map the polynomial multiplications in SABER on crossbars and show that analog dot-products can yield a $1.7-32.5\times$ performance and energy efficiency improvement, compared to recent hardware proposals. This initial design combines the innovations in multiple state-of-the-art works -- the algorithm in SABER and the memristive acceleration principles proposed in ISAAC (for deep neural network acceleration).  We then identify the bottlenecks in this initial design and introduce several additional techniques to improve its efficiency.
These techniques are synergistic and especially benefit from SABER's power-of-two modulo operation.
First, we show that some of the software techniques used in SABER, that are effective on CPU platforms, are unhelpful in crossbar-based accelerators.  Relying on simpler algorithms further improves our efficiencies by $1.3-3.6\times$.
Second, we exploit the nature of SABER's computations to stagger the operations in crossbars and share a few variable precision ADCs, resulting in up to $1.8\times$ higher efficiency.
Third, to further reduce ADC pressure, we propose a simple analog Shift-and-Add technique, which results in a $1.3-6.3\times$ increase in the efficiency. Overall, our designs achieve $3-15\times$ higher efficiency over initial design, and $3-51\times$ higher than prior work.
Finally, analog operations are more error-prone; we show that the approximate nature of module LWR calculations can mask most of these errors. We also show that such hardware-induced errors do not violate the security guarantees of the algorithm. 

\end{abstract}

\section{Introduction}\label{sec:01-introduction}
The recent emergence of several quantum computing systems -- IBM's Q
system~\cite{ibmq20}, Intel's Tangle Lake~\cite{hsu18}, Google's Bristlecone~\cite{kelly18}, and
IonQ~\cite{wang18} -- has increased the likelihood that integer factorization and discrete logarithm will be
tractable in the near future, thus rendering several modern-day cryptographic
primitives obsolete~\cite{bernsteinlange17,wolchover15}.  This has spurred interest in alternative
cryptographic assumptions that cannot be easily solved by known algorithms
(Shor~\cite{shor99} and Grover~\cite{grover97}) on quantum systems.  In the past several years,
NIST has solicited and short-listed a number of quantum-resistant algorithms~\cite{alagic19}.
A number of mathematical approaches are being considered for such post-quantum cryptography (PQC),
including lattice-based, multivariate, hash-based, isogeny-based, and code-based
cryptography~\cite{bernsteinlange17,alagic19,nejatollahidutt19}.  Of these, lattice-based cryptography (LBC)
is a front-runner because of its high efficiency and security on
several metrics~\cite{bernsteinlange17,nejatollahidutt19}.  Research on hardware acceleration of these PQC algorithms is therefore time-critical as it can steer the field towards algorithms that are
more amenable to hardware acceleration.  This research is also timely because,
even though the algorithms continue to evolve, there is convergence on the
basic primitives and computations that will be included in most algorithm
variants. 
Several companies, including Google, Microsoft, Digicert, and Thales, are already testing the impact of deploying PQC~\cite{kris19}.

Modern infrastructures are built on cloud-based deployments that secure their data in transport using a mix of symmetric and asymmetric key encryption. 
Typically, a private key is established using a handshake protocol based on asymmetric key encryption. Following this, data in transit is secured using this private key.
However, asymmetric key encryption schemes like RSA/ECC are vulnerable to quantum attacks; these may be replaced by their PQC counterparts that have been proposed recently. 
As we will describe shortly, popular LBC schemes based on Learning With Errors (LWE)~\cite{regev09} (and its variants) rely on large matrix and vector operations that place a
significant burden on the hardware.  Recent
efforts~\cite{basusoni19,howemoore16,agrawalbu19,nejatollahidutt18,leeseo21,imranalmeida21,zhuzhu21,ghoshmera22} implement
LBC algorithms on FPGAs/GPUs/ASICs and report latencies of tens to hundreds of micro-seconds.
Classic cryptographic functions can be typically executed in \textit{micro-seconds} on modern hardware~\cite{mclvormcLoone03,singhkhan16}; for instance Intel QuickAssist technology executes RSA decrypt operation in 5 us using a specialized crypto-accelerator~\cite{IntelQuickAssist}.  The metrics for LBC therefore fall well short of the demands of modern deployments. 
Several contemporary HTTPS applications require key-establishment between a web-server and client (usually million requests every second), using asymmetric key cryptographic primitives. 
PQC algorithms, with software implementations in hundreds of micro-seconds~\cite{ribeiroda}, would \textit{increase service time by at least an order of magnitude}.
In order to replace the pre-quantum cryptographic algorithms used in this scenario, such as RSA~\cite{rivestshamir78} and Diffie–Hellman~\cite{diffiehellman76}, PQC algorithms will have to consume significantly lower latency and energy.
While higher parallelism can
reduce latency, it can incur a significant cost in terms of area, data
movement, and energy.  It is therefore vital to explore transformational
new hardware technologies that can improve all relevant efficiency metrics by
orders of magnitude.  Without these advancements, the next generation of privacy-preserving
deployments -- homomorphic encryption that further scales up the computation or
medical IoT devices that further scale down the resource constraints -- will be
out of reach.

This paper explores a promising technology -- analog computing in
resistive memories -- as the foundation for new architectures that efficiently
execute a range of algorithms relevant to LBC.  While
such technologies have been used before~\cite{shafieenag16,chili16} to implement computations in deep neural networks, we show that LBC
offers new opportunities to further improve the efficiency of this compute-in-memory approach.
Although in-memory analog operations are efficient at performing dot-products, the conversion of a high-resolution current into a digital value is expensive.  This analog to digital conversion (ADC) is the primary overhead that must be kept in check.  A second challenge with analog circuits is that they can be noise-prone.  We overcome these challenges by exploiting the following opportunities in PQC algorithm variants: (i) the computations can tolerate some noise, (ii) they perform power-of-2 modulo operations.  

\begin{figure}[h!]
\centering
\includegraphics[width=1.0\linewidth]{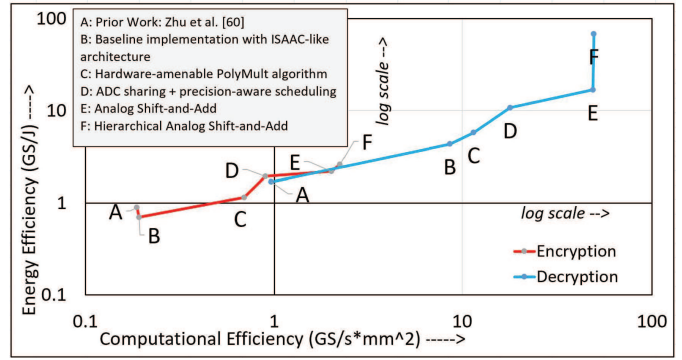}
\caption{Summary of techniques introduced in this paper and their impact.}
\label{ressummary}
\end{figure}

We focus on SABER, one of the NIST finalists that relies on a Module-LWR algorithm.  Adopting the best practices from prior work in DNN acceleration, we create an initial design that out-performs prior FPGA-based PQC acceleration efforts. We then analyze this initial design and with hardware-software co-design, further improve decryption compute density by $9.5\times$ ($11.8\times$ for encryption) and energy efficiency by $15.7\times$ ($3.7\times$ for encryption).  These improvements are achieved by incorporating accelerator-amenable software techniques for polynomial multiplication, staggered computation to reduce ADC pressure, and a novel analog shift-and-add technique. The summary/impact of each novel technique is illustrated in Figure~\ref{ressummary}. The techniques introduced in this work can be extended to all PQC algorithms that base operations on polynomial math, like Digital Signatures and Homomorphic Encryption. 

We make the following key contributions:
\begin{itemize}
    \item We construct a basic in-memory architecture for PQC encryption and decryption algorithms, leveraging prior best practices~\cite{shafieenag16,chili16}. This design serves as the baseline for this study and achieves higher efficiency than recent PQC accelerators. 
    \item We demonstrate software techniques like decomposing polynomial degree and smart scheduling to increase ADC sharing, that improves energy efficiency by 2.4-2.7$\times$ and computational efficiency by 4.7$\times$, relative to our baseline. 
    \item We propose a novel technique to perform write-free in-analog shift-and-add operations using crossbars, allowing us to trade-off cell programming with ADC complexity. A design space exploration yields an up to $2.7\times$ and $6.3\times$ increase in computational and energy efficiency, respectively.
    \item We show that lattice-based schemes can tolerate small amounts of error in computations, introduced by analog device/circuit variations.
    \item Overall, XCRYPT (Xbar based accelerator for post-quantum CRYPTography) achieves server deployment with decryption latency of 0.08 us and client deployment with encryption latency of 4 us, with an overall chip area of just 0.04 and 0.3 $mm^2$ respectively.
\end{itemize}


\section{LBC Background} \label{sec:02-background}

\subsection{Learning With Errors and Its Variants}
\label{lbcbasics}

Given the popularity of lattice-based approaches, this paper
focuses on LBC.  A lattice is a set of points in $n$-dimensional space.  Each
point is a linear integer combination of a set of basis vectors that define
the lattice.  Assuming that the basis vectors are large, given a point, it is
computationally expensive (non-polynomial time) to determine the nearest
lattice point.  Intuitively, the high dimension and the many possible
linear combinations of large vectors contribute to making this a difficult problem.
The difficulty of the Closest Vector Problem~\cite{goldreichmicciancio99} and
other related problems (Shortest Vector Problem, Shortest Independent Vector
Problem) lay the foundation of many cryptographic primitives, such as
public key encryption, digital signature and homomorphic encryption~\cite{gentry09}.

The hardness of some LBC schemes are based on the hardness of Learning with Errors (LWE~\cite{regev09}) problem. 
The Standard LWE problem states that, given a randomly chosen $n$-dimensional vector $\vec{s}$ of integers modulo $q$, 
it is hard to recover $\vec{s}$ from $m \ge n$ \textit{approximate} random linear equations on $\vec{s}$. In other words, we start with a secret value $\vec{s}$; we multiply $\vec{s}$ with $m$ random vectors,
and add a small error $e_i$, to generate $\vec{b}$ (public key).
Instantiations of LWE schemes differ in the parameter space ($n, q, m$), and distributions used to sample $\vec{s}, e$. 

In Ring-LWE~\cite{lyubashevskypeikert10}, integer samples are replaced with polynomial samples.
The main computation here is polynomial multiplication (\textit{PolyMult}), which can be implemented efficiently using Number Theoretic Transform (NTT), which is a variant of FFT. However, Ring-LWE may compromise security, compared to Standard LWE~\cite{eliaslauter15,castryckiliashenko16}. In order to retain the increased efficiency from Ring-LWE while providing higher security guarantees, Module-LWE~\cite{bosducas18} was proposed.  It uses an $l$-dimensional vector of polynomials. Its efficiency can be further improved by using a rounding operation instead of padding with randomly drawn errors.
This variant is called Learning With Rounding (Module-LWR)~\cite{alwenkrenn13}.
{\em Two of the NIST round 3 candidates use module structures: Kyber~\cite{bosducas18} uses Module-LWE, and SABER~\cite{danverskarmakar18} uses Module-LWR.}

\subsection{SABER} \label{subsec:SABER_background}

NIST has solicited algorithms for Key Encapsulation Mechanisms (KEMs) and Digital Signatures.  SABER~\cite{danverskarmakar18} is a finalist among the KEMs. 
The SABER KEM is composed of three algorithms - key generation, encryption, and decryption.

Key generation determines a matrix $\mat{A}$ of polynomials using a pseudo-random number generator based on SHAKE-128~\cite{dworkin15}.
A secret vector $\vec{s}$ of polynomials is generated by sampling from a centered binomial distribution. The public key consists of the matrix seed and the rounded product $\vec{b}=\mat{A}^T\vec{s}$. 

Encryption generates a new secret $\vec{s'}$, and adds the message $m$ (a polynomial with coefficients $\in \{0,1\}$) to the inner product of public key $\vec{b}$ and $\vec{s'}$, forming the first part of ciphertext. The second part hides the encrypting secret by rounding the product $\mat{A}\vec{s'}$. 

The third algorithm is Decryption, which uses the secret key $\vec{s}$ to extract the message encoded in the two parts of ciphertext.


Overall, SABER performs 24 PolyMults, 14 Poly Modulo, and 8 Poly Rounding operations -- each PolyMult does $256^2$ integer products and Modulo/Rounding are applied per coefficient.
SABER adds a few constants in its calculations so that rounding can be replaced with simple bit shifts.

In module-lattice based cryptography, the performance of PolyMult plays a key role in the overall performance~\cite{kannwischerrijneveld19}. 
In our experiments, we observed that PolyMult kernels consume $>90\%$ of the execution time. 
When implementing PolyMult, NTT has the asymptotically fastest time complexity of $O(n\log{n})$, but it requires $q$ to be a prime, which in turn leads to non-trivial complexity for modulo operations.
On the other hand, SABER chooses a power-of-two $q$, which speeds up the modulo (by simply dropping the MSBs).  Since NTT is not an option, SABER uses the Toom-Cook-4 algorithm~\cite{cookaanderaa69} to reduce each degree-256 PolyMult to 7 degree-64 PolyMults, and then further reduces them to degree-32 using the Karatsuba algorithm once.
This choice (along with AVX2 support) brings the contribution of PolyMult to 57\%, and SABER outperforms implementations (like Kyber) that employ a faster NTT-based polynomial multiplier. SABER demonstrates computations in tens of micro-seconds using AVX2 support~\cite{danverskarmakar18}. Various HW/SW co-designs of SABER have reported lower latencies while utilizing fewer resources~\cite{dangfarahmand19,roybasso20,meraturan20}, compared to FPGA/ASIC implementations of other lattice-based schemes~\cite{banerjeeukyab19,howeoder18}. 
Additionally, SABER has among the lowest ciphertext overhead among the candidates. We compare the 3 lattice-based PQC candidates and their NIST security level 1 KEM implementations in Table~\ref{table:NISTcomparison}. Given its favorable properties in terms of speed and lower hardware/ciphertext complexity,  we choose SABER as the target for our hardware acceleration.

In addition to polynomial multiplication, generating pseudorandom numbers using SHAKE-128 is also a costly operation.
It is well known that Keccak-core, which is at the heart of SHAKE-128, is very efficient on hardware platforms~\cite{roybasso20}. We observed Keccak functions to contribute $<30\%$ of the execution time, also validated by \cite{roybasso20}.  Further, the Keccak-core can be accelerated~\cite{team19} and pipelined with other arithmetic operations.
In this work, we therefore focus on the main bottleneck of LBC, the polynomial multiplication.

\begin{table}[]
\caption{Comparison of NIST Round 3 LBC schemes}
\label{table:NISTcomparison}
\begin{tabular}{|c|c|ccc|}
\hline
\multirow{2}{*}{PQC scheme} & \multirow{2}{*}{\begin{tabular}[c]{@{}c@{}}Ciphertext/\\ Plaintext ratio\end{tabular}} & \multicolumn{3}{c|}{Haswell Cycles} \\ \cline{3-5} 
 &  & \multicolumn{1}{c|}{KeyGen} & \multicolumn{1}{c|}{Encaps} & Decaps \\ \hline
Kyber-768~\cite{bosducas18} & 37.00 & \multicolumn{1}{c|}{85K} & \multicolumn{1}{c|}{112K} & 108K \\ \hline
SABER~\cite{danverskarmakar18} & 34.00 & \multicolumn{1}{c|}{101K} & \multicolumn{1}{c|}{125K} & 129K \\ \hline
NTRU~\cite{hoffsteinpipher98} & 40.03 & \multicolumn{1}{c|}{307K} & \multicolumn{1}{c|}{48K} & 67K \\ \hline
\end{tabular}
\normalsize
\end{table}

\subsection{Existing PQC implementations}
Efforts are
already underway to implement PQC algorithms in
hardware~\cite{nejatollahidutt19}. Most of the early efforts have focused on FPGA
implementations~\cite{basusoni19,howemoore16,agrawalbu19,nejatollahidutt18}.  We discuss some of the salient efforts here as motivation.

A comprehensive discussion of PQC hardware/software efforts~\cite{gottertfeller12,poppelmannguneysu12,poppelmannguneysu13,aysupatterson13,royvercauteren14,cousinsgolusky14,wangchen14,wanghu13,dubai15,gyorfioctavian13,dubai16a,dubai16b,banerjeeukyab19,nejatollahicammarota19,nejatollahishahhosseini20,farahmanddang19,bassoroy20,merakarmakar20,imranalmeida21,zhuzhu21,ghoshmera22,leeseo21} can be found in the
survey paper by Nejatollahi et al.~\cite{nejatollahidutt19}.  They note that most of the
hardware overhead can be attributed to multiplication units; the Gaussian
Sampler unit also incurs non-trivial hardware cost.  Basu et
al.~\cite{basusoni19} use high-level synthesis to generate FPGA circuits for PQC algorithms.  
Reported latencies are in the micro-seconds range, with
a significant fraction of the hardware overhead in the polynomial multiplier
(even after using the $O(n\log n)$ NTT algorithm). \cite{basusoni19,farahmanddang19} evaluate various NIST candidates on their software/hardware codesign implementations. \cite{merakarmakar20} focus on algorithmic optimizations for polynomial multiplication. Note that efficient FPGA
implementations require a specific set of techniques that focus on reducing the
bottleneck component -- look-up tables (LUTs). Recent works have highlighted using GPUs~\cite{leeseo21} or their ASIC designs~\cite{imranalmeida21,zhuzhu21,ghoshmera22}.
We compare our basic design against many of these solutions in Section~\ref{subsec:03-02-SABER_on_Xbar}.

\subsection{Impact of KEM Deployments}
In modern cloud services, millions of new clients request services every minute, each requiring a symmetric key establishment using KEM. Quantum-safe KEM routines like SABER have significantly higher complexity than traditional routines, thus lowering quality-of-service. While this paper focuses on KEMs, the benefits seen in KEM acceleration will also apply to other applications based on PolyMult. Quantum-secure digital signature schemes like CRYSTALS-DILITHIUM~\cite{ducaslepoint18} (NIST Finalist) are based on PolyMult. 
LBC also forms the basis for  Homomorphic Encryption (HE). Popular HE schemes like FV/BFV~\cite{fanvercauteren12,brakerski12} (implemented in several libraries like FV-NFLlib from CryptoExperts~\cite{cryptoexperts16} and SEAL from Microsoft~\cite{seal16}), GSW~\cite{gentrysahai13}, CKKS~\cite{cheonkim17}, and TFHE~\cite{chillottigama20} encode information in polynomials. Our work can be extended to HE-applications, where polynomial math is also the key bottleneck. Moreover, contemporary HE schemes work with a noise budget; when the budget is consumed, decryption and encryption of ciphertext is required using SABER or SABER-like schemes (the target application in this paper). Cheetah~\cite{reagenchoi21} demonstrated Secure Machine Learning as a service using HE, and required decryption/encryption for each output neuron after every layer, thus increasing the contribution of the KEM scheme. A deeper analysis of HE and digital signature deployments is left as future work. 

\section{Accelerator Background}\label{pimbasics}
In the past decade, industry and academia have made significant investments in
resistive memory technologies~\cite{burrbreitwisch10,vontobelrobinett09,leeipek09,izraelevitzyang19}.  A resistive memory cell uses its
resistance value to store information.  
Resistive memory arrays have several advantages -- very high
density, non-volatility, and competitive read latency/energy.  They also have a
primary challenge -- writes are expensive because they consume more energy/time
and cause device wearout.  After years of research, the first generations of
commercial resistive memory products have emerged in the last few years~\cite{izraelevitzyang19}.
More recently, multiple groups~\cite{hustrachan16,chili16,shafieenag16,kvatinskybelousov14,songqian17} have identified that a resistive memory array can be configured to perform analog operations.  Such {\em processing-in-memory} technologies have the potential for high parallelism and low data movement.

\begin{figure}
\centering
\includegraphics[width=1.0\linewidth]{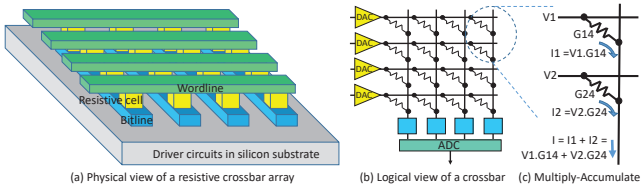}
\caption{Resistive crossbar (Xbar) circuit and how it performs an analog dot-product operation~\cite{shafieenag16,hustrachan16}}
\label{fig:dpe}
\end{figure}


A resistive memory array is implemented as a {\em crossbar} -- a grid of
cells, as shown in Figure~\ref{fig:dpe}a.  Sandwiched
between the X-dimension wires (wordlines) and the Y-dimension wires (bitlines)
are the resistive cells (a material with programmable resistance).
Figure~\ref{fig:dpe}a shows the physical layout of a small crossbar.
Figure~\ref{fig:dpe}b shows the
logical representation of the crossbar, while Figure~\ref{fig:dpe}c zooms into
the operation in one bitline.
If voltages
$V_1$ and $V_2$ are applied to the first two wordlines, current is injected
into each bitline, proportional to the conductance of each cell in those rows.
The current in the last bitline is
$V_1\times G_{14} + V_2\times G_{24}$, where $G_{ij}$ represents the
conductance of cell $\{i,j\}$.  Thus, the current in each bitline is an analog
representation of the dot-product of two vectors -- the voltage vector applied
to the wordlines and the conductance vector pre-programmed into a column of
cells.

Further, the wordline voltage is broadcast to all the columns; each column
performs an independent and parallel dot-product on the same voltage
vector, but using a different conductance vector.  The basic Kirchoff's Law equation is being exploited to design an analog
vector-matrix multiplication circuit that yields a vector of output bitline
currents in a single step when a vector of input voltages is applied to the
wordlines.  
Programming the cells is a
time-consuming process.
Performing the vector-matrix multiplication is equivalent to a crossbar read
operation followed by analog sensing, an operation that is much faster (order
of $10^{-7}$ seconds).
Prior works~\cite{hustrachan16,shafieenag16,chili16} have shown that this analog matrix-vector multiplication unit
can achieve orders of magnitude reduction in energy per operation (depending
on the type of analog sensing), compared to an equivalent digital circuit.  This is partly because the computation
is ``in-situ'', i.e., the matrix is an operand that doesn't have to move and
partly because complex multiplications and additions are being performed by
exploiting natural phenomena (Kirchoff's Law).

The analog signal emerging from each bitline has to be converted into a
digital value.  Analog to digital conversion (ADC) circuits consume
significant power that grows exponentially with ADC resolution.  Managing
ADC resolution and power is a key challenge in exploiting the
capabilities of this emerging technology.
\cite{kvatinskybelousov14,chili16,nejatollahigupta20} have explored other circuits as well.

Recent works~\cite{shafieenag16,chili16,songqian17,feinbergwang18,choutang19,narayananshafiee17} have employed these analog resistive circuits as the
central component in accelerators for DNNs, which can tolerate small noise in computations. 
This has inspired several projects, including commercial startup efforts~\cite{fick18,crossbar20,memryx20,syntiant20}.
This paper builds on the insight from these prior architectures and
explores if cryptographic applications can exploit this emerging new
technology.  In particular, the approximate nature of LBC makes it a potential
``killer application'' for analog resistive circuits.


Recent in-situ DNN accelerators have highlighted the significant overhead of converting analog signals to digital. ISAAC~\cite{shafieenag16} spreads the computation over time and area to reduce ADC precision to 8-bits, hence reducing ADC overhead to 58\% of the power and 31\% of the silicon area. PRIME~\cite{chili16} uses sense amplifiers (SAs) instead of conventional ADCs to reduce area. However, since a SA is only capable of resolving one bit at a time, it takes $2^8$ cycles to convert an 8-bit signal. CASCADE~\cite{choutang19} writes the output signal to another memristor crossbar in order to perform shift and adds before converting the signal to digital. This reduces the number of ADC calls by almost $25\times$. However, CASCADE is only effective with low cell programming overhead, which is unlikely~\cite{zahoorazni20,shenzhao20} given their need for 6-bit resolution and low write voltage. 
\section{LBC on Memristor XBar Arrays}\label{sec:03-PQC_on_Xbar}

Most prior work on LBC acceleration has focused on FPGA implementations~\cite{roybasso20,basusoni19,farahmanddang19,bassoroy20,meraturan20,dangfarahmand19} and achieved latencies and energy in the 5us-1ms and 10uJ-10mJ range, respectively. Recent literature has shown that in-RRAM computation can improve the efficiency of dot-product heavy DNN applications~\cite{shafieenag16,chili16,songqian17,nagbalasubramonian18,choutang19}.  We now describe how memristive crossbars can be an ideal substrate for dense and energy-efficient calculations for LBC. 

{\em We first create a strong baseline by adapting the ISAAC architecture~\cite{shafieenag16}, which was designed for DNNs.  Next, we make the case that SABER's PolyMult algorithm (Toom Cook-4 + Karatsuba) is not always well suited for memristor crossbars, and explore different multiplication algorithms. Then, we show techniques, enabled by SABER's power-of-2 modulo, that reduce the overheads of ADCs. In the next section, we extend this design to allow shift and adds in analog, further lowering ADC requirements.} 

Similar to AES encryption engines in today’s processors, the XCRYPT accelerator will also be transparent to programmers and will be invoked by hardware when the software needs to send secure messages. Note that the proposed accelerator is based on memristors (RRAM), which are CMOS compatible and can be integrated on the same die as the processor.

\begin{figure}
    \centering
    \includegraphics[width=\linewidth]{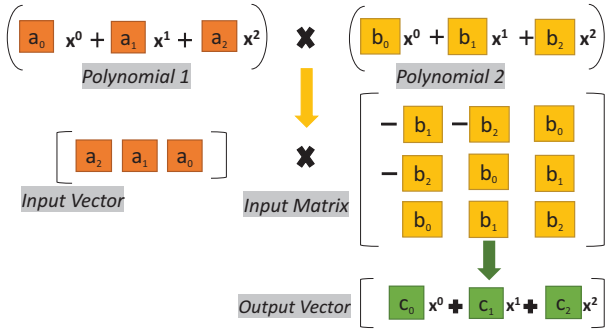}
    \caption{Representing modular polynomial multiplication as vector-matrix multiplication (shown for degree=3).}
    \label{fig:poly_mult_asVMmult}
\end{figure}

\subsection{Modular PolyMult as Vector-Matrix Multiplication}\label{subsec:03-01-poly_mult_as_VMmult}
At its core, almost all lattice-based candidate schemes in NIST PQC do modular polynomial multiplication. \textit{Modular} here means that the resultant polynomial of multiplication of two $n$ degree polynomials $p_1, p_2 \in R_q = \Z_q[x]/(x^n+1)$ would have its  $x^i (i \ge n)$ terms divided by $-x^n$ to keep the degree $n$. This is what dividing polynomial $\Z_q[x]$ by $(x^n+1)$ represents.
Polynomial multiplication requires each coefficient of $p_1$ to be multiplied with each coefficient of $p_2$. Consider degree-3 polynomials, $p_1 = a_0x^0+a_1x^1+a_2x^2$ and $p_2 = b_0x^0+b_1x^1+b_2x^2$; $p_1 \times p_2 = (a_0b_0-a_1b_2-a_2b_1)x^0 + (a_0b_1+a_1b_0-a_2b_2)x^1 + (a_0b_2+a_1b_1+a_2b_0)x^2$. As seen in Fig~\ref{fig:poly_mult_asVMmult}, this can be viewed as vector $(a_2, a_1, a_0)$ being multiplied by a matrix with columns $(-b_1, -b_2, b_0), (-b_2, b_0, b_1), (b_0, b_1, b_2)$. The column for last term $x^2$ is $(b_0, b_1, b_2)$, while the rest are downward single shifted versions of it with a negative value at the column head.

\begin{table}[]
\caption{XCRYPT-Schoolbook (X-SB) parameters, at 32nm.}
\label{table:params}
\begin{tabular}{|ll|l|l|}
\hline
\multicolumn{1}{|l|}{\textbf{Component}} & \textbf{Count} & \textbf{Power (uW)} & \textbf{Area (um$^2$)} \\ \hline
\multicolumn{1}{|l|}{XBar (128x128)} & 1 & 300 & 25 \\ \hline
\multicolumn{1}{|l|}{DAC 1bit} & 128 & 3.9 & 0.16 \\ \hline
\multicolumn{1}{|l|}{S+H 6bit} & 128 & 0.007 & 0.029 \\ \hline
\multicolumn{1}{|l|}{ADC 6bit} & 16 & 945 & 435 \\ \hline
\multicolumn{1}{|l|}{ADC 7bit} & 1 & 1365 & 628.33 \\ \hline
\multicolumn{2}{|l|}{\textbf{Total (1 array)}} & \textbf{123.13} & \textbf{7737.557} \\ \hline
\multicolumn{2}{|l|}{\textbf{\begin{tabular}[c]{@{}l@{}}X-SB = 1 Encryption + \\ 1 Decryption Tile = 2 x\\ (48 arrays)+(IR+OR+SA)\end{tabular}}} & \textbf{11.92 mW} & \textbf{0.743 mm$^2$} \\ \hline
\end{tabular}
\end{table}

\subsection{Methodology}\label{subsec:03-02-Methodology}
Before we describe our proposed accelerator XCRYPT, we first state our methodology. This makes it convenient to discuss XCRYPT design choices with supporting results. We leverage many of the primitives introduced in the ISAAC architecture~\cite{shafieenag16} and adopt an evaluation methodology very similar to that work. The energy and area model for memristor crossbar arrays, including Shift-and-Add Crossbars, is based on that of Hu et al.~\cite{hustrachan16}. The RRAM cell model is derived from \cite{zahoorazni20}, with 25~ns write latency and 0.1~pJ/cell/bit write energy, and NVSim~\cite{dongxu12} is used to extract array level numbers. Read latency is determined by the ADC readout as RC delay of the crossbar is typically sub-ns~\cite{hustrachan16,hugraves18,jizhang19}. 
Since RRAM is an emerging technology, there are numerous RRAM device parameters accepted within the research community targeting different implementations. We have confirmed that our proposed techniques are independent of these parameter values and yield significant improvements for a wide range of parameters. 
The read energy of an RRAM cell is four orders of magnitude lower than write energy. Area and energy for shift-and-add, sample-and-hold, and 1-bit DAC circuits are adapted from ISAAC~\cite{shafieenag16}. We have considered an energy and area efficient adaptive ADC that can handle one giga-samples per second~\cite{choobell16}. The adaptive ADC has three major components - Charge-based DAC, Comparator, and Controller. To arrive at power/area for different ADC bit precision, we scaled power/area of charge-based DAC exponentially, and the rest linearly.
Detailed parameters of our initial design (X-SB, described in next subsection) are listed in Table~\ref{table:params}. Note that we propose various versions of XCRYPT with varying XBar/ADC sizes, leading to various parameters.
We also model CASCADE components based on parameters mentioned by Chou et al.~\cite{choutang19}. We modify the code for NIST contestant SABER~\cite{danverskarmakar18} to execute various XCRYPT design features. We further extend the code to model RRAM cell variance in order to calculate the noise tolerance of the implementation, more of which is described in Section~\ref{subsec:04-04-noise_analysis}. Simulations are done for a million randomly generated key pairs which are used to encrypt and decrypt 32B plaintext, and report the failure rate of decryption. We consider 2 key metrics to evaluate \AccName{} efficiency:
\begin{enumerate}
    \item Computational Efficiency (CE): Number of 1-bit plaintext/ciphertext operations per second per mm$^2$ of area \\$(Gbits/s \times mm^2)$. 
    \item Energy Efficiency (EE): Number of 1-bit plaintext/ciphertext operations per 1J of energy $(Gbits/J)$.
\end{enumerate}

We consider the SABER variant that has post-quantum security level similar to AES-192.
In SABER, decryption 
requires vector-vector multiplication 
while encryption 
requires both vector-vector 
and matrix-vector 
multiplications. A vector contains $l=3$ polynomials of degree $n=256$, while a matrix consists of $l^2=9$ polynomials. Therefore, a vector-vector multiplication does $3$ PolyMults, followed by their addition, while matrix-vector multiplication does $9$ PolyMults to obtain a vector of 3 polynomials. SABER's parameters define the coefficients of polynomials in matrix-vector multiplication to be $\log{q}=13$ bits. For the 
multiplication calls, the input polynomials are rounded to have $\log{p}=10$ bits. In all PolyMults, one of the polynomials is either $\vec{s}$ or $\vec{s'}$, which are both secrets sampled from a centered binomial distribution 
such that each coefficient requires 4 bits.  

\subsection{Mapping SABER to Memristor Crossbars}\label{subsec:03-02-SABER_on_Xbar}

In this sub-section, we first create a strong crossbar-based baseline to execute SABER, following the principles of the ISAAC design for DNNs~\cite{shafieenag16}.  Because RRAM writes are a bottleneck in this design, we introduce a couple techniques to alleviate this overhead.  This baseline is then compared against prior work in PQC acceleration before we further address the bottlenecks in the baseline.

In a cloud service platform, several client requests are serviced every second. To guarantee secure communication, the client and server establish a symmetric key, through SABER's PKE (RSA in the non-quantum realm). A server, therefore, generates its key once per client. Using the key pair, it establishes connections with clients by repeatedly invoking the encryption and decryption algorithms that are the focus of this paper. 

\begin{figure}
    \centering
    \includegraphics[width=1\linewidth]{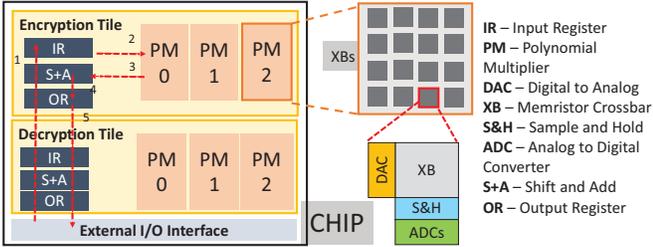}
    \caption{XCRYPT-SchoolBook or X-SB architecture. Arrows denote the path for Encryption.}
    \label{fig:architecture}
\end{figure}
\begin{figure}
    \centering
    \includegraphics[width=1\linewidth]{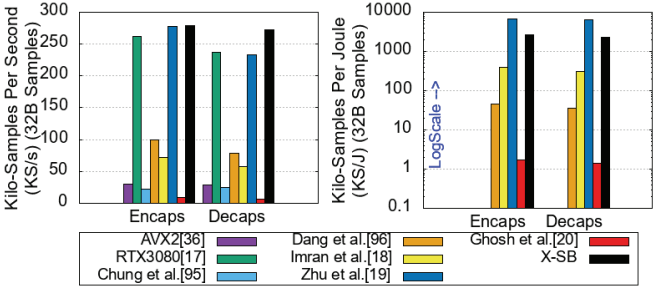}
    \caption{SABER on memristor crossbars v/s published implementations of SABER. \cite{imranalmeida21,zhuzhu21,ghoshmera22} are ASICs.}
    \label{fig:comparison_relatedwork}
\end{figure}

We use a methodology very similar to that for ISAAC~\cite{shafieenag16} to map the required computations to crossbars. Since secret key $\vec{s}$ remains constant for much of the server's runtime, we designate $\vec{s}$ as the operand to be encoded in crossbar cells.  We need 72 128$\times$128 1-bit crossbars to store key $\vec{s}$ and its shifted versions (as needed by PolyMult).  A tile in our design (shown in Figure~\ref{fig:architecture}) has multiple crossbars and can perform 3 PolyMults, as required by a vector-vector multiplication.  Using ISAAC's flip-encoding scheme, each column produces a 6-bit current.  For our work, we use an area-efficient ADC~\cite{choobell16} that can convert samples at a frequency of 1~GSps. To further reduce area overhead, we share 1 ADC across 8 columns of a crossbar, which results in crossbar read cycle time of 8~ns.

\noindent {\bf Handling RRAM Writes.}
As illustrated in Figure~\ref{fig:architecture}, we design 2 different tiles - encryption and decryption. This is because we target two different deployments - a cloud server and a client device. In a typical handshake protocol, the server and client establish a private key using asymmetric key encryption (like SABER in the post-quantum world). During the protocol, client encrypts using the server's public key while the server decrypts using its secret key. Hence, we design XCRYPT encryption/decryption tiles for client/server respectively. 

Programming the memristor cells is expensive, in terms of energy, performance, and endurance. Typically, memristors-based cells have a budget of $10^{12}$ writes before the cells malfunction.
Since the server only periodically changes its private key (for freshness), the number of writes to the memristor is low enough to sustain a reasonable lifetime. For instance, considering a conservative assumption of a private key update every second, the accelerator lifetime exceeds 30K years. On the client side, every new connection demands a new memristor write. However, the typical number of connections established by a client is small. Even with $10^5$ connection setups per day, the client accelerator lifetime will exceed 27K years.

\noindent {\bf Comparison to Baselines.}
We have thus defined an initial XCRYPT design modeled after ISAAC principles and an efficient write policy.
We designate this architecture as \textit{XCRYPT-SchoolBook} or \textit{X-SB} because it uses the schoolbook algorithm for polynomial multiplication.
We compare this initial design with existing implementations of SABER in Figure~\ref{fig:comparison_relatedwork} to show that this basic design out-performs prior work on LBC acceleration. 
Most prior work report Encapsulation/Decapsulation numbers. These functions have encryption/decryption as their underlying kernels -- Encaps calls encryption once, while Decaps calls encryption and decryption once. For a fair comparison, we compare Encaps/Decaps in Figure~\ref{fig:comparison_relatedwork}. However, throughout the paper, we report results for the \textit{underlying independent} kernels, encryption and decryption.

The CPU implementation results are reported from the SABER paper~\cite{danverskarmakar18}, running on an Intel Haswell machine at 3.1~GHz, with AVX2 support. 
Lee et al.~\cite{leeseo21} evaluates capabilities of GPUs for running PQC (results shown for RTX3080 in figure).
Chung et al.~\cite{chunghwang21} changes SABER's parameters to enable the NTT transformation and benefit from a fast 16bit NTT multiplier in hardware. 
Dang et al.~\cite{dangmohajerani} also proposed a new design of Saber based on NTT, demonstrated on FPGA.
Imran et al.~\cite{imranalmeida21} performs a design space exploration of ASIC with smart pipelining and logic sharing between SABER kernels. Their baseline architecture is ported from an FPGA proposal for SABER hardware~\cite{roybasso20}. 
Zhu et al.~\cite{zhuzhu21} use an 8-level hierarchical Karatsuba framework for PolyMult and a task scheduler that reduces resource utilization by up to 90\%. Their post-layout chip is approximately the same size as our X-SB processor, which gives a fair comparison. 
Ghosh et al.~\cite{ghoshmera22} showcase the first Silicon verified ASIC implementation for Saber.
{\em Compared to existing works, RRAM can achieve 1.17$\times$ higher throughput budget, as seen in Figure~\ref{fig:comparison_relatedwork}.}
While state-of-the-art ASIC~\cite{zhuzhu21} achieves 2.4$\times$ better energy efficiency than X-SB, we demonstrate techniques in this paper that enable XCRYPT to outperform ~\cite{zhuzhu21} by 2 orders of magnitude.
Memristor crossbars are efficient at performing multiply-accumulate using Kirchhoff's law. Analog dot products are performed at sub-ns RC delay, and are readout at the peripheral's digital conversion frequency. As observed in previous works~\cite{shafieenag16,choutang19,hustrachan16}, this ADC circuit is the bottleneck, contributing 70-90\% of total energy.

CryptoPIM~\cite{nejatollahigupta20} is the first paper to propose PIM acceleration for lattice-based schemes, demonstrating throughput improvement over existing FPGA work, but significantly higher latency. That work underestimates the programming overhead of RRAM cells. Each operation in their proposal requires writing to RRAM cells, with an assumed latency of 1.1ns. However, realistically each RRAM cell takes 25ns to program~\cite{choutang19}. Even if we assume parallel write drivers to simultaneously write cells in an entire row/column, we estimate that CryptoPIM’s polynomial multiplication takes 5 orders of magnitude higher latency than our baseline X-SB design, at the cost of 13$\times$ higher energy. Furthermore, RRAM-based devices are vulnerable to non-idealities, which is overlooked in CryptoPIM. CiM~\cite{reistakeshita20} also proposes PIM solutions for PolyMult, targeted at Homomorphic Evaluation (HE), which typically has a much larger parameter set than public-key cryptography. They therefore focus on an SRAM substrate with a ComputeCache implementation~\cite{agajeloka17}. On the other hand, XCRYPT implements a smaller parameter SABER on area-efficient RRAM devices, and targets the ADC bottleneck.

\subsection{Impact of Polynomial Multiplication Algorithms}\label{subsec:03-03-poly_algos}

\begin{figure}
    \centering
    \includegraphics[width=1\linewidth]{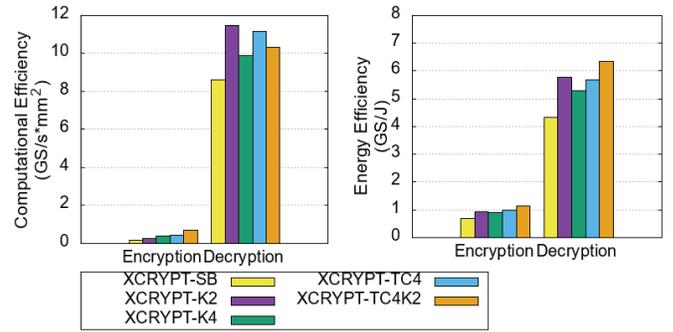}
    \caption{CE and EE for SABER on memristor crossbars, evaluating various polynomial multiplication algorithms.}
    \label{fig:Algos}
\end{figure}

\begin{figure}
    \centering
    \includegraphics[width=1\linewidth]{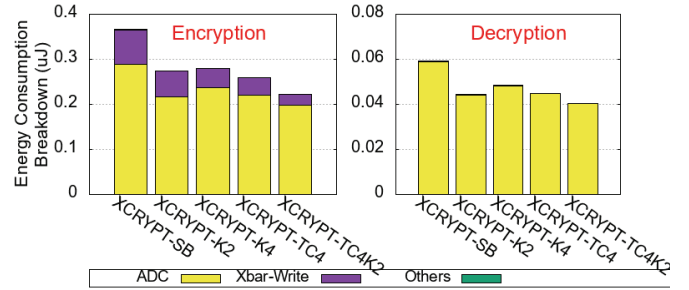}
    \caption{Energy breakdown of XCRYPT, for various multiplication algorithms, for encryption and decryption.}
    \label{fig:Algos_EnergyBreakdown}
\end{figure}

In this sub-section, we explore different multiplication algorithms to identify the implementation that is most amenable to crossbar acceleration. Similar to matrix multiplication, polynomial multiplication also benefits from various lower complexity algorithms~\cite{nejatollahidutt19}. Figure~\ref{fig:Algos} quantifies CE and EE for these algorithms. We start with the standard $O(n^2)$-complexity Schoolbook algorithm, designated \textit{X-SB}, for multiplying 256-degree polynomials. Karatsuba's algorithm ($O(n^{1.58})$-complexity) breaks down a 256-degree multiplication to 3 128-degree PolyMults, reducing the number of required 128$\times$128 crossbars from 32 to 24. This proportionally decreases ADC and crossbar write energy, as depicted by \textit{X-K2} in Figure~\ref{fig:Algos_EnergyBreakdown}.
Note that decryption doesn't require crossbar writes because of a constant secret $\vec{s}$, which is why almost all of the decryption energy in Figure~\ref{fig:Algos_EnergyBreakdown} is contributed by the ADC. Encryption also has a lower CE than decryption as it requires more crossbars, has more input cycles, and requires a long initial crossbar programming step (90\% of end-to-end encryption latency). 

\textit{X-K4} further reduces the 3 128-degree multiplications to 9 64-degree PolyMults, which requires a smaller 64$\times$64 crossbar. While X-K4 reduces the degree and the ADC samples per crossbar, it also increases the total number of PolyMults. Due to this, the total number of ADC samples, and the crossbar writes increase, which is why X-K4 results in lower improvements 
than X-K2 for decryption. Note that X-K4 maps to a smaller 64$\times$64 crossbar, enabling a lower write energy contribution than X-K2, despite more crossbar writes. 

ToomCook-4 ($O(n^{1.4})$-complexity) reduces a 256-degree multiplication to 7 64-degree PolyMults. While ToomCook-4 is asymptotically faster than K2, it creates more polynomials that require increased ADC samples and crossbar requirements, which explains the worse behavior for \textit{X-TC4}, relative to \textit{X-K2}, for decryption. However, TC4 performs well during encryption as the benefits from a smaller crossbar (lower write latency and energy) outweigh the higher crossbar count requirements. Software implementations of SABER~\cite{danverskarmakar18} use ToomCook-4 + Karatsuba-2 (labeled as \textit{X-TC4K2} in the figures) to do 21 32-degree PolyMults, which performs best for encryption due to its small crossbar write latency/energy. For encryption, X-TC4K2 improves CE and EE over X-SB by 3.6$\times$ and 1.6$\times$, respectively. 
Lowering the degree beyond 128 does not result in higher efficiency for decryption where no writes happen. {\em Thus, the ideal SABER algorithm on a crossbar accelerator varies, and we choose X-K2 for decryption and X-TC4K2 for encryption, for the rest of the paper.}

\begin{figure}
    \centering
    \includegraphics[width=\linewidth]{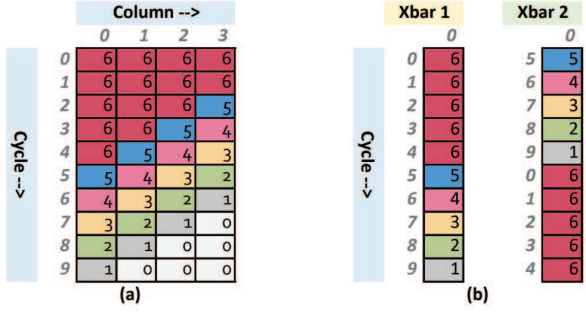}
    \caption{(a) Number of bits that contribute to the final coefficient. (b) Reordering computations in crossbars to enable sharing of maximum precision ADCs.}
    \label{fig:ADC_Sharing}
\end{figure}
\subsection{Using Modulo to Reduce ADC Overheads}\label{subsec:03-04-ADCShare}
The \textit{X-K2} implementation of SABER on memristor crossbars is primarily constrained by ADC overhead.  The ADC consumes 90\% of the area, and 78\% of the energy during encryption. Rest of the energy is attributed to crossbar writes. Similar ADC overheads are reported by previous studies for DNN applications as well~\cite{shafieenag16,choutang19,hustrachan16}. In the 128$\times$128 crossbar, with ISAAC's flip encoding scheme, a 6-bit precision ADC is required to convert the analog value to digital~\cite{shafieenag16}. Since the computations are spread across cells in a row and across cycles, the 6-bit ADC results have to be aggregated after appropriate shifts. SABER's parameters define that the coefficients of the output polynomial must go through modulo $2^{10}$ or $2^{13}$, which keeps the coefficients at 10 or 13 bits. Thus, given the modulo operation, not all bits from all 6-bit ADC samples contribute to the final output. {\em Unlike DNN computations, where most significant bits carry the most information, some most significant bits in our computations are ineffectual.} For instance, all 6 bits from cycle 0's LSB column are needed as they add to the LSB 6 bits of the output coefficient. However, the output of cycle 9's LSB column are added to the same coefficient after shifting left 9 times, i.e., only the LSB of the sampled ADC value will contribute to the final 10-bit result. In Figure~\ref{fig:ADC_Sharing}(a), we illustrate the number of relevant bits while doing decryption. Each coefficient of secret $\vec{s}$ is 4 bits, which is stored across 4 cells in a row, with LSB stored in column 0. Therefore, the outputs of 4 columns have to be added after appropriate shifts. Furthermore, values across cycles are also shifted-and-added. This is shown in Algorithm~\ref{algo:pseudocode_for_coeff}. As seen from the figure, {\em the number of bits and hence, the ADC precision varies depending upon the value of (cycle+column).} For vector-vector multiplications, we require full precision (6 bits) only for (cycle+column) $\le 4$.
\begin{algorithm}
\small
\SetAlgoLined
\caption{Pseudo-code for polynomial multiplication, per coefficient, using crossbar}\label{algo:pseudocode_for_coeff}
coeff = 0;\\
for(cycle=0; cycle$<$10; ++cycle)\\
\quad for(column=0; column$<$4; ++column)\\
\quad\quad coeff += bitline\_value[cycle][column]$\ll$(cycle+column);\\
coeff = (coeff)$\mod{2^{10}}$\\
\normalsize
\end{algorithm}

\begin{figure}
    \centering
    \includegraphics[width=\linewidth]{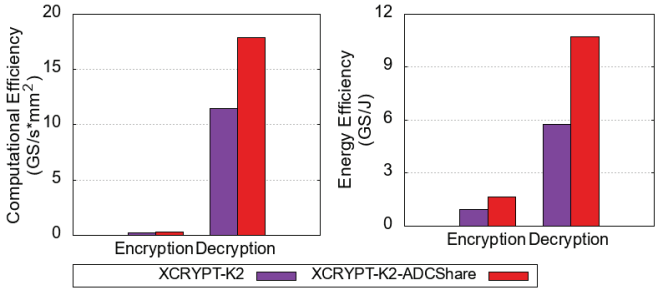}
    \caption{Improvement by exploiting low precision computations and ADC sharing by reordering computations.}
    \label{fig:ADCShare}
\end{figure}

{\em We take advantage of this flexibility by reordering another crossbar's computations in such a manner that at any given cycle, at most 1 crossbar is producing an output with maximum precision of 6.} This is illustrated in Figure~\ref{fig:ADC_Sharing}(b), which depicts the column 0 output precision requirements for 2 crossbars. The computations for the second crossbar are reordered such that the 5$^{th}$ gets executed in the 0$^{th}$ cycle. This staggering ensures that only one crossbar produces a 6-bit output in a given cycle. By sharing 2 ADCs, one 6-bit and the other 5-bit among the crossbars, we lower the ADC overheads relative to the baseline with two 6-bit ADCs.  Each of the ADCs handles half the workload.  This concept can be further applied to lower energy while slightly increasing area.  The 5-bit ADC workload can be split across a 5-bit ADC and a 4-bit ADC.  The 4-bit ADC handles nearly 90\% of this split workload, thus saving energy.  The area overhead of an extra ADC can be reduced by sharing the 5-bit ADC across 10 crossbars (since the 5-bit ADC is assigned a small fraction of the conversions).  Since ADC overheads grow exponentially with its precision, this technique of leveraging shared, lower precision ADCs allows us to improve EE by 1.8$\times$ and CE by 1.5$\times$ over X-K2 in decryption, as seen in Figure~\ref{fig:ADCShare}. Since this technique doesn't reduce the number of crossbar writes, benefits in encryption are lower - 1.7$\times$ in EE and 1.08$\times$ in CE. 

This highlights the importance of choosing modulo as power of 2, which leads to the above {\em ADCShare} technique and significant benefits for SABER. {\em Other FFT-based PQC schemes, which have modulo over large primes, require computations of all the intermediate values before modulo is applied.} 
\cite{reistakeshita20,danverskarmakar18} have demonstrated benefits in modifying schemes to have power-of-2 modulo without affecting security; so the \textit{ADCShare} technique will apply to a broad range of lattice-based PQC algorithms. 
The \textit{ADCShare} technique does not apply to DNNs because dropping intermediate bits can impact the final result.
SABER, on the other hand, reduces the output by performing modulo with $2^t$, which means that any intermediate result that contributes to the $(t+1)^{th}$ bit is unnecessary.

\section{Shift-and-Add Crossbars (SACs)}\label{sec:03-SACs}

PolyMult with crossbars generates many intermediate ADC readout values, that are later shifted-and-added to obtain the final output polynomial. For instance, each PolyMult in decryption generates 4 values per cycle, for 10 cycles, which are appropriately shifted-and-added to obtain one output coefficient value (Algorithm~\ref{algo:pseudocode_for_coeff}). 
Intermediate analog value readouts are expensive since they are done using ADCs. In this section, we explore the possibility of performing the shift-and-add operation in analog to delay the ADC readout.

\subsection{Existing In-analog Shift-and-Add Implementation}\label{subsec:04-01-CASCADE}
CASCADE~\cite{choutang19} proposed in-analog shift-and-add of intermediate values by writing the output of a crossbar's columns to another crossbar called Buffer RRAM crossbar. In a given cycle, the column outputs are written in adjacent cells of a row of the Buffer crossbar, while values across cycles are written to different rows with appropriate shifts. With this mapping, a simple readout of the Buffer crossbar performs shift-and-add of all intermediate values. By performing shift-and-add in analog, CASCADE delays the ADC readout to a single final value. 
However, CASCADE assumes a lower RRAM cell write energy overhead than that reported in recent literature. 
While multiple write drivers allow parallel cell programming, the write latency is expected to be about 25~ns.
RRAM cells also have 4 orders of magnitude higher write energy (0.1~pJ/cell/bit~\cite{zahoorazni20}) than read energy~\cite{zahoorazni20,chenli18,hajriaziza19,shenzhao20}. 
We next discuss an alternative analog shift-and-add technique that is more efficient than CASCADE.

\subsection{Write-Free In-Analog Shift-and-Add using SACs}\label{subsec:04-02-SAC}

We propose a novel technique to perform shift-and-add in analog. 
We introduce Shift-and-Add Crossbars (SACs) - tiny single column crossbars whose cells are pre-programmed to hold successive powers of 2. The intuition behind SAC is that given an input vector, when passed through SAC, individual values are multiplied by cell values and aggregated.  
In practice, up to 6-bit precision RRAM cells have been demonstrated. Therefore, SACs, with highest multiplier factor of 1$\ll$5, can only add inputs with a maximum of 5 shifts. However, this can be overcome with hierarchical deployment of SACs. 

\begin{figure}
    \centering
    \includegraphics[width=0.9\linewidth]{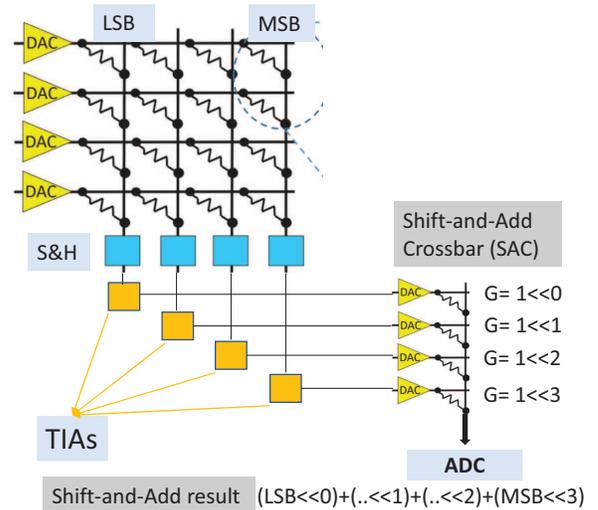}
    \caption{Performing Shift-and-Add on outputs of 4 columns using Shift-and-Add Crossbars (SACs).}
    \label{fig:SAC_1XBar}
\end{figure}

\begin{figure}
    \centering
    \includegraphics[width=1\linewidth]{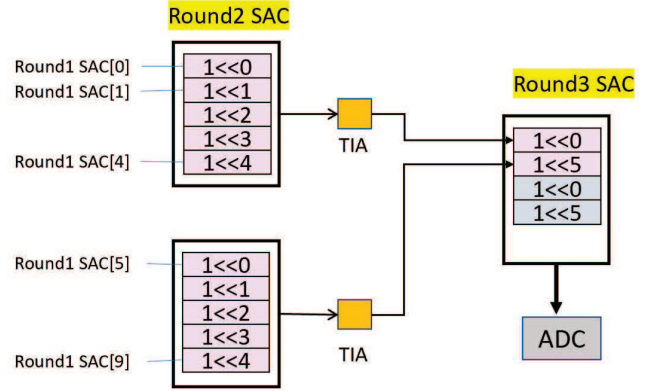}
    \caption{Hierarchical-SAC: In-analog accumulation across all cycles for each output polynomial coefficient, for SAC-All.}
    \label{fig:HierarchicalSAC}
\vspace{-8pt}
\end{figure}

\begin{figure*}[!ht]
    \centering
        \subfloat[Encryption]{\includegraphics[width=\columnwidth]{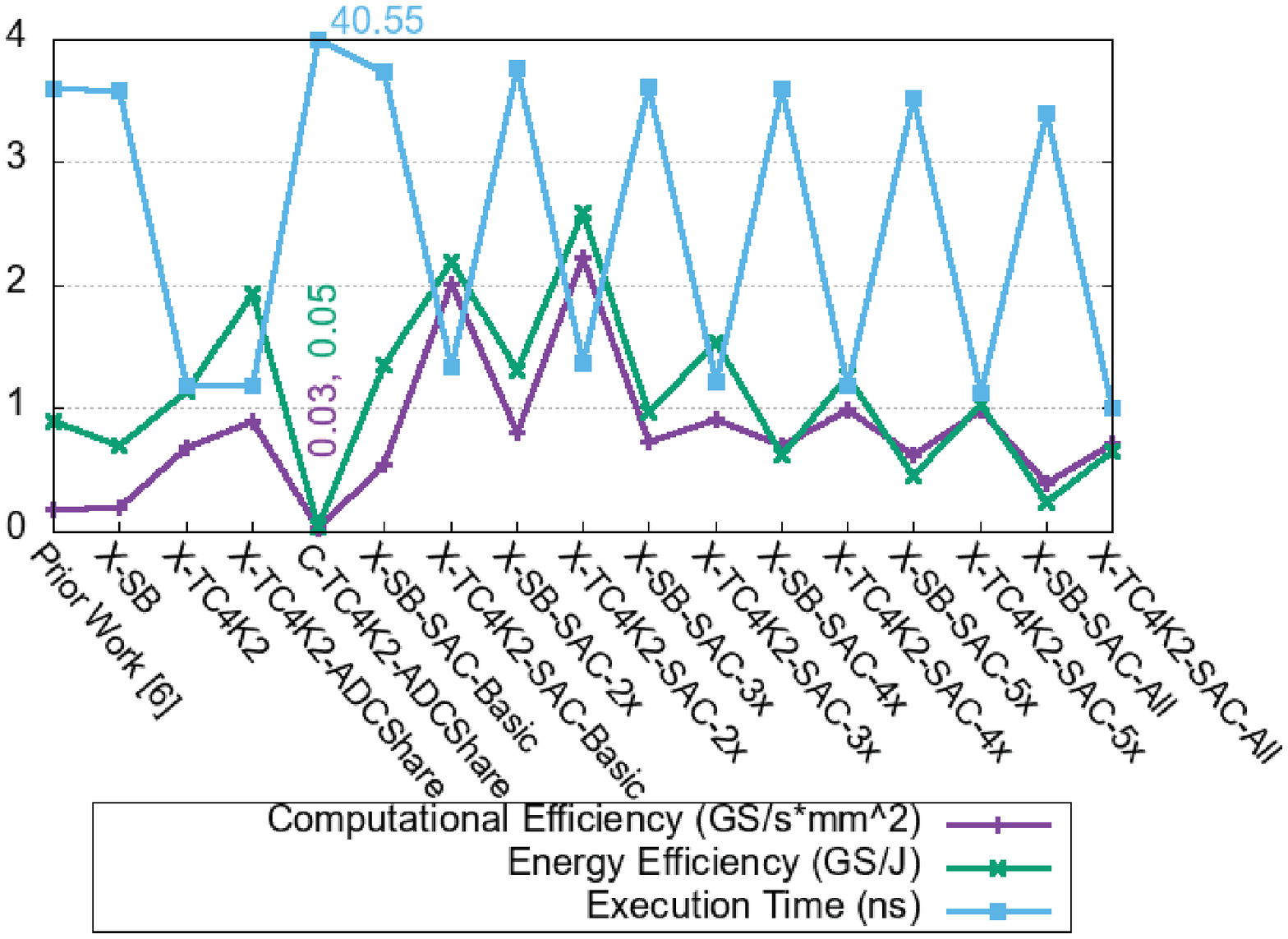}\label{fig:SAC_Enc}}
        \subfloat[Decryption]{\includegraphics[width=\columnwidth]{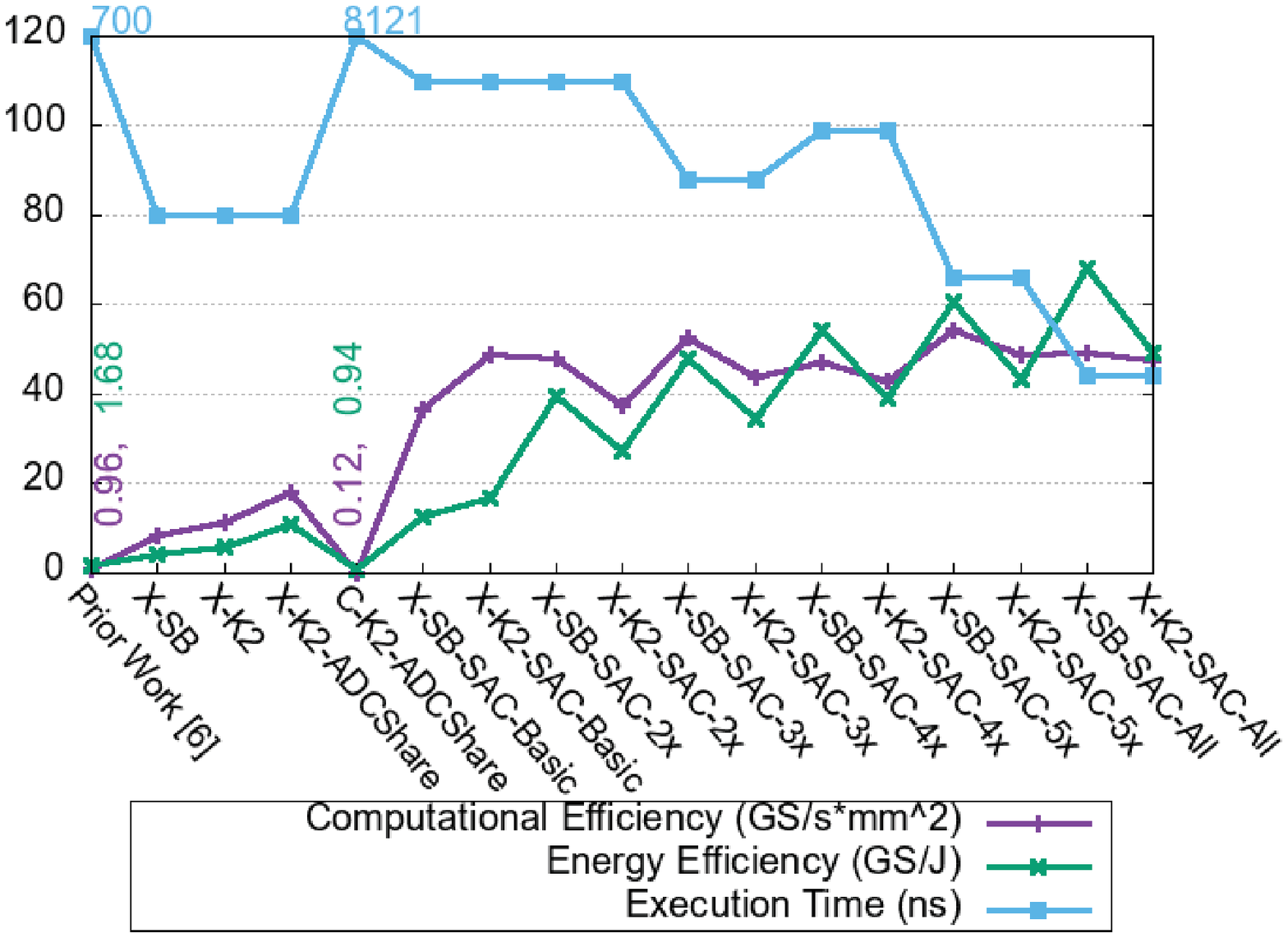}\label{fig:SAC_Dec}}
    \caption{Comparison of various implementations of XCRYPT designs. \textit{X-}, \textit{C-} represents XCRYPT, CASCADE respectively.}
    \label{fig:Result_SACs}
    \vspace{-8pt}
\end{figure*}

\begin{figure*}[!ht]
    \centering
        \subfloat[Encryption]{\includegraphics[width=\columnwidth]{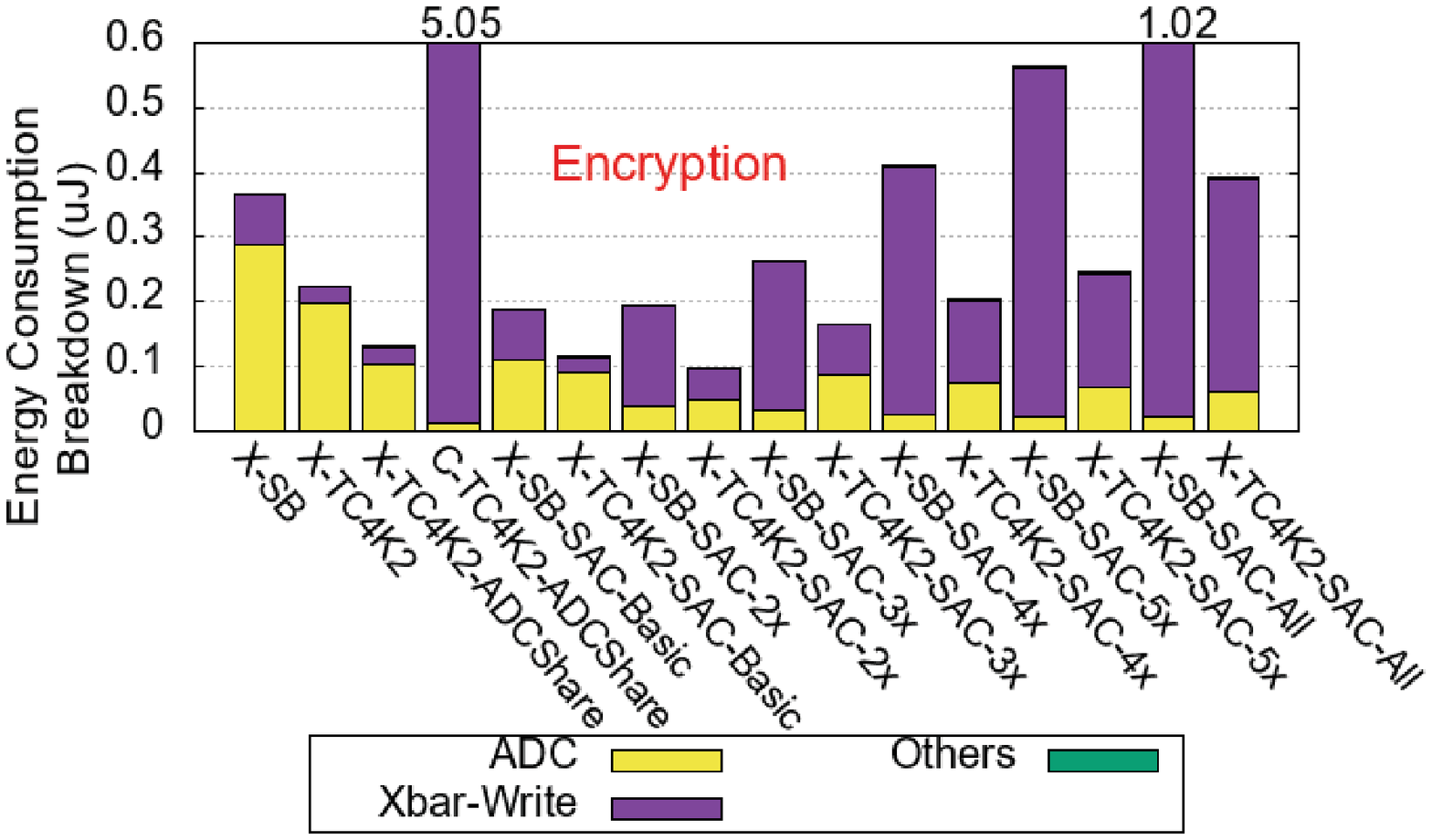}\label{fig:SAC_Enc_EnergyBreakdown}}
        \subfloat[Decryption]{\includegraphics[width=\columnwidth]{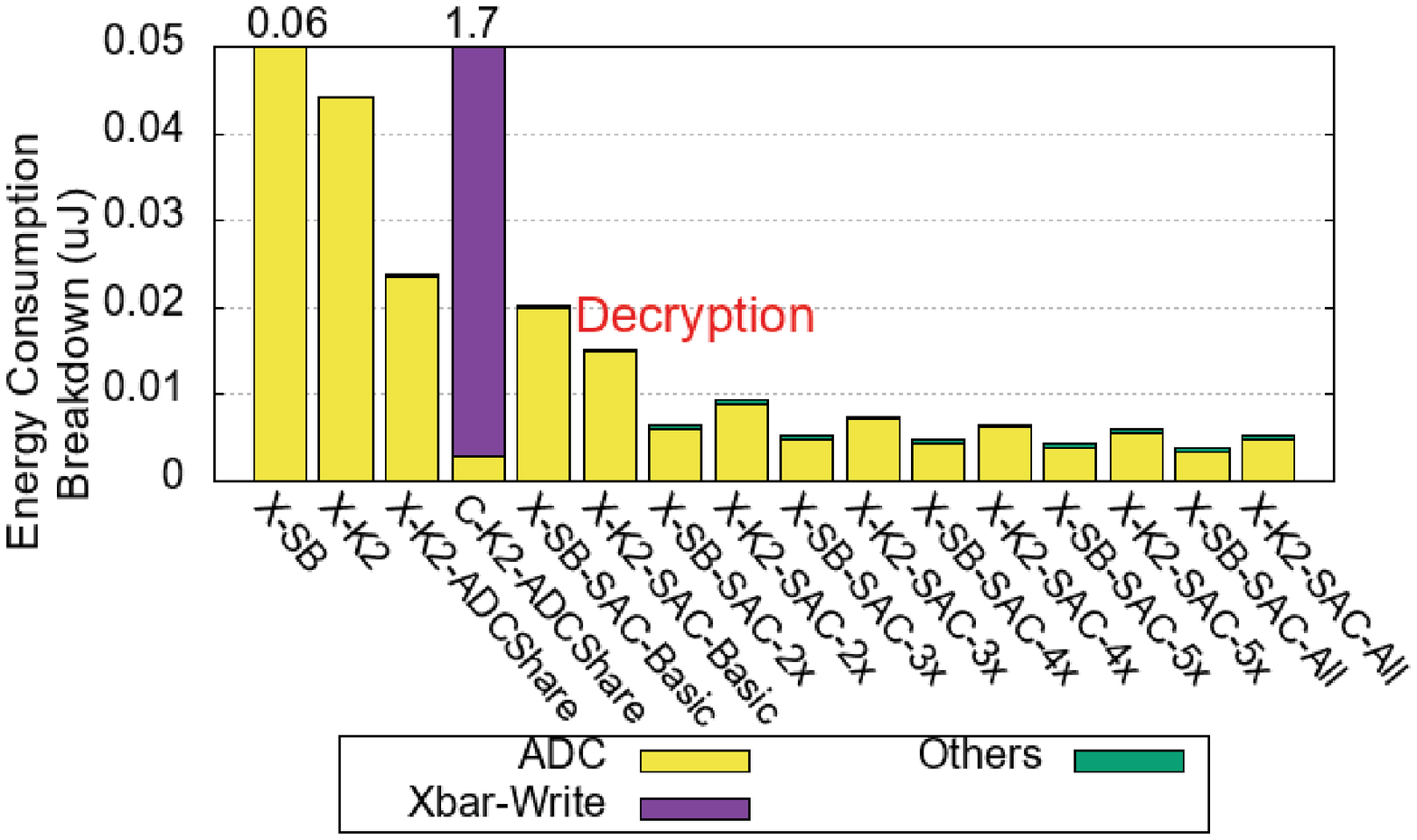}\label{fig:SAC_Dec_EnergyBreakdown}}
    \caption{Energy Breakdown of various XCRYPT designs. \textit{X-}, \textit{C-} represents XCRYPT, CASCADE respectively.}
    \label{fig:SAC_EnergyBreakdown}
    \vspace{-3pt}
\end{figure*}

We first start by using SACs within a cycle. Since the secret $\vec{s}$ is written to 4 1-bit cells in a row, the output of 4 columns must be added with appropriate shifts, every cycle. In the baseline implementation, this addition happens in digital, after ADC readout. In our work, we propose using a single SAC to perform these shift-and-adds, as demonstrated in Figure~\ref{fig:SAC_1XBar}. In order to feed a crossbar's output current to SAC's DAC, it must be first converted to a proportional voltage signal, which is usually done using TransImpedance Amplifiers (TIAs).
We use a fast (11~ns sense+transfer time) TIA circuit proposed by CASCADE~\cite{choutang19}.
Note that SAC's output is the shift-and-add result of 4 6-bit values, resulting in a 10-bit value. Therefore, a more expensive 10-bit precision ADC is required. However, by delaying ADC readout, it converts 4$\times$ fewer samples and allows more sharing within the crossbar. Moreover, as the cycles proceed, fewer bits from the accumulated value contribute to the final output coefficient, as described in Section~\ref{subsec:03-04-ADCShare}. This enables flexibility to increase ADC sharing, as only a single 10-bit sample is generated across the whole dot product. On the other hand, in a non-SAC implementation, 4$\times$ 6-bit samples are generated during many cycles, as seen from Figure~\ref{fig:ADC_Sharing}. The SAC significantly lowers the number of samples, increasing the effectiveness of the ADC sharing and smart scheduling described in Section~\ref{subsec:03-04-ADCShare}. We refer to this design as \textit{SAC-Basic}.  SAC-Basic is a synergistic technique that exploits known analog circuits (like TIA), the flexibility offered by SABER's modulo operation, and resource sharing through smart scheduling on crossbars.

\subsection{In-analog Accumulation Across Cycles}\label{subsec:04-03-SAC_across_cycles}
In the previous subsection, we showed how to accumulate column outputs within a cycle in analog, delaying the ADC readout. Since input values are also streamed 1-bit per cycle, ADC outputs across different cycles must also be shifted-and-added once in digital. 
Since our design does not write the cycle output to a crossbar (like CASCADE), it cannot readily accumulate across cycles. However, we can overcome this roadblock by doing multiple cycles in parallel.


\noindent {\bf Accumulate across cycles: }
SAC requires inputs to be fed simultaneously across different rows in order to shift-and-add them. Therefore, to add outputs from 2 cycles, their computations need to be done in parallel. 
This requires 2$\times$ compute/storage resources but reduces ADC samples by 2$\times$ -- a reasonable trade-off as ADC accounts for a majority of the energy/area. 

We label the per-crossbar single-cycle SAC as \textit{Round1-SAC}. Unlike SAC-Basic, 
Round1-SAC's outputs are redirected to another SAC using TIA. 
This SAC is shared across multiple crossbars that contribute to the same output coefficient, and is termed as \textit{Round2-SAC}. This 2-cycle accumulated column value is finally read out with an ADC at the end of Round2-SAC. We term this design as \textit{SAC-2x} - accumulating 2 cycles before readout. While SAC-2x reduces ADC overheads, it increases the number of crossbars that run in parallel, in turn increasing the overall crossbar write costs. Therefore, we perform a design space exploration evaluating this trade-off while increasing the number of cycles that are accumulated in-analog. 
At the extreme end, SAC-All performs in-analog accumulation of all columns over all cycles per output polynomial coefficient, similar to CASCADE, and hence generates only 1 ADC sample per output coefficient.

\noindent {\bf Hierarchical SACs: }
Since the RRAM cell has a max resolution of 6 bits, larger computations are performed hierarchically, thus extending to Round3 and Round4 SACs, as shown in Figure~\ref{fig:HierarchicalSAC}. Note that SAC techniques are applicable to other lattice-based PQC schemes and algorithms like HE as well.


\subsection{Results}

\noindent {\bf SAC-Basic Results: }
We compare our basic XCRYPT design with ADC Sharing techniques, CASCADE with similar techniques, and our novel design with SAC adding column values (\textit{SAC-Basic}) within a cycle in Figure~\ref{fig:Result_SACs}. Cycle time for the SAC design is determined by the TIA's sense+transfer time (=11ns). In a cycle, inputs are streamed to the first crossbar, sensed by TIA, and then sent as inputs to SAC. In the second cycle, TIA's outputs are streamed through SAC, producing the final value, which is sampled using ADC. Due to increase in the number of cycles and cycle time, end to end latency increases, relative to basic XCRYPT. However, by delaying ADC readout, SAC-Basic achieves 2.7$\times$ (2.4$\times$ for encryption) higher CE and 1.5$\times$ (1.1$\times$) higher EE, over X-K2-ADCShare (X-TC4K2-ADCShare). 
On the other hand, CASCADE (\textit{C-K2-ADCShare}) performs worse than the basic design, 
highlighting the overheads of performing crossbar writes every cycle. Other applications like DNNs can also benefit from the SAC design.

\noindent {\bf Parallel SAC results: }
We also compare various SAC-* designs in Figure~\ref{fig:Result_SACs}. 
As we increase the parallelism to reduce ADC samples, the bottleneck shifts from ADCs to crossbar writes.
We plot energy breakdown results in Figure~\ref{fig:SAC_EnergyBreakdown}. SAC designs offer a trade-off between ADC and crossbar write energy. From SAC-Basic to SAC-All, the bottleneck shifts from ADC (77\% of total energy) to crossbar writes (83\% of total energy). The benefits from fewer ADC samples cannot outweigh the energy overheads of more crossbar writes, which is why SAC-All's EE decreased by 66\% from X-TC4K2-ADCShare, for encryption. Meanwhile, since decryption doesn't need to write to the crossbar (after being written once on boot), increasing the level of parallelism increases CE and EE.



We also observe that {\em the use of these circuit techniques changes the software algorithm that is most amenable to acceleration}, e.g., Schoolbook out-performs Karatsuba in some cases. 
Overall, the best decryption design (\textit{X-SB-SAC-All}) yields 2.7$\times$ and 6.3$\times$ increase in CE and EE, respectively, over X-K2-ADCShare. For encryption, these improvements for \textit{X-TC4K2-SAC-2x} are 2.5$\times$ and 1.3$\times$ over X-TC4K2-ADCShare.

\noindent {\bf Results Summary:}
Overall, the best decryption design (\textit{X-SB-SAC-All}) yields 4.5$\times$ and 6.3$\times$ increase in CE and EE, respectively, over X-K2-ADCShare. For encryption, these improvements for \textit{X-TC4K2-SAC-2x} are 2.5$\times$ and 1.3$\times$ over X-TC4K2-ADCShare. Compared to state-of-the-art ASIC~\cite{zhuzhu21}, XCRYPT shows 3-51$\times$ higher efficiencies with 2.6-16$\times$ speedup. A single tile can perform $0.7M/22.7M$ encryptions/decryptions per second with a $0.09/0.07~mm^2$ area budget. Looking at the broader KEM operations, Encaps can be performed in 1.3 us with an area budget of 0.08 $mm^2$ and 0.16 uJ, while Decaps can be done within 1.3 us with 0.16 $mm^2$ area and 0.17 uJ energy.

\begin{figure}
    \centering
    \includegraphics[width=\linewidth]{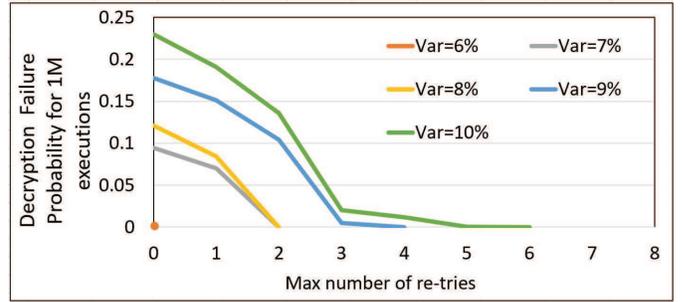}
    \caption{Number of re-tries needed to achieve failure-free decryption for 1M SABER calls, varying Cell Variance.}
    \label{fig:noise_analysis}
\end{figure}

\section{Noise Analysis}\label{subsec:04-04-noise_analysis}

Due to non-ideal device behaviors and circuit issues, in-RRAM computations are vulnerable to errors~\cite{yeochu19,chenshih14,hustrachan16,hugraves18,chakrabortyali20,liuhu13,charanhazra20,jainraghunathan19}. However, we argue first that although the noise generated during PolyMults can affect the correctness of decryption, it \textbf{does not leak information} about the secret key. In the first step of SABER's decryption algorithm, instead of computing $v$ as $\langle \vec{b}', \vec{s} \rangle$ (ignoring the rounding operation), $v$ is actually set as $\langle \vec{b}', \vec{s} \rangle + e$, where $e$ is the noise caused by the cell variation. Since the decryption fails, the sum of errors in the computation plus $e$ exceeds the rounding parameter, where $e$ accounts for the majority of the sum and is much larger than the norm of the secret key. Then, the noisy value $\langle \vec{b}', \vec{s} \rangle + e$ is ``essentially independent'' of the secret key $\vec{s}$. This is sometimes referred to as “noise flooding”~\cite{goldwasserkalai10} since the noise $e$ ``floods” the value related to the secret key and minimizes its effect. Under CPA-secure SABER, decryption based attacks are not applicable, as mentioned in SABER specs~\cite{saberspec}.  

Many approaches use calibration signals~\cite{hustrachan16,hugraves18,chakrabortyali20,liuhu13} or fault-aware mapping/encoding~\cite{yeochu19,chenshih14}, or a combination of the two~\cite{charanhazra20,jainraghunathan19}, mostly for machine learning applications. Hu et al.~\cite{hugraves18} have demonstrated the use of TIA for simple linear calibration in order to achieve low noise in practical deployment. We simulate our SAC-All design, with cell variance ranging up to 10\%, and plot the failure probability for 1 million executions of SABER's public-key algorithms in Figure~\ref{fig:noise_analysis}. 
Cell variance is the variance of the distribution function that introduces non-ideality to the output. Specifically, a 5\% cell variance means that its output would be within 5\% range of the ideal cell \textit{output current}. Our crossbar error is modeled at the level of every single multiply-accumulate operation (individual RRAM cell) as we sample from a probability distribution function for variation. 
The noise in the TIA circuit is conservatively assumed to be 2\%. 
We have modeled non-idealities occurring from cell programming and variation for the SAC circuitry, with a constant cell variation, which might not hold true when a large current passes through the device~\cite{chakrabortyali20}. However, unlike the conventional Shift-and-Add where the errors can be amplified due to bit-shifts in digital, the errors accumulated in SAC are over current and would only reflect during ADC readout. Which is why readout, by converting a range of currents to digital values, results in lower error than conventional shift-and-add. Single cell failures can affect the scheme’s failure probability. However, such issues can be easily identified and remapped~\cite{yeochu19,chenshih14}.

To detect failed decryption, data is accompanied by its CRC~\cite{petersonbrown61}. Upon failure, we re-try that computation. In Figure~\ref{fig:noise_analysis}, we show the decryption failure probability as a function of the cell variance and the maximum number of allowed re-tries. As observed from the graph, our accelerator achieves failure-free decryption with cell variance of 5-6\%. Beyond that, variation in cells triggers the bitline current to jump over an interval, resulting in incorrect coefficients and hence, failed decryption with up to 0.23 probability. With a few re-tries, these errors can also be corrected.  In practice, 5\% cell variance is reasonable~\cite{xiali17,hustrachan16,liuhu13,choiyang14}, and can be further lowered with recent calibration techniques. Furthermore, with state-of-the-art fabrication that has higher device yield, lower wire resistance, and that operates at lower conductances, the noise impact can further reduce.

\section{Conclusion}
This work evaluates the use of memristor crossbars for accelerating lattice-based post quantum cryptography (PQC). We show that even a simple implementation of SABER, a front-runner PQC candidate for NIST Round-3, performs faster than existing hardware proposals for SABER. By exploiting SABER's algorithmic properties, e.g., its power-of-2 modulo operations, we can further boost the accelerator's efficiency.  We identify polynomial multiplication as the key operation in lattice-based schemes, and show that crossbar-based designs might not benefit from some of the existing software techniques for multiplication. We propose SABER-specific variable precision ADCs, which, along with computation reordering, allows high levels of ADC sharing.
To further reduce ADC overheads, we propose simple analog Shift-and-Add techniques. Overall, our proposed accelerator achieves 12-51$\times$/3-40$\times$ higher computational/energy efficiency than state-of-the-art ASIC.
Finally, we modify SABER's NIST submission code to simulate noisy crossbars, and present a noise analysis for XCRYPT design points.

\bibliographystyle{unsrt}
\bibliography{main}

\end{document}